\documentclass[acmsmall,screen,nonacm]{acmart}

\setcopyright{acmlicensed}
\copyrightyear{2018}
\acmYear{2018}
\acmDOI{XXXXXXX.XXXXXXX}

\acmConference[Conference acronym 'XX]{Make sure to enter the correct
  conference title from your rights confirmation emai}{June 03--05,
  2018}{Woodstock, NY}
\acmISBN{978-1-4503-XXXX-X/18/06}

\usepackage[shortlabels]{enumitem}
\usepackage{mathtools}
\usepackage{subcaption}
\usepackage{mathpartir}
\usepackage{cleveref}
\usepackage{listings}
\usepackage{xcolor}
\usepackage{wrapfig}
\usepackage{threeparttable}
\usepackage{multirow}
\usepackage{tikz}

\crefname{theorem}{Thm.}{Thms.}
\crefname{lemma}{Lem.}{Lemmas}
\crefname{corollary}{Cor.}{Cors.}
\crefname{figure}{Fig.}{Figs.}
\crefname{definition}{Defn.}{Defns.}
\crefname{table}{Tab.}{Tabs.}
\crefname{appendix}{Appendix}{Appendices}
\crefformat{section}{\S#2#1#3}
\crefmultiformat{section}{\S#2#1#3}{ and~\S#2#1#3}{, \S#2#1#3}{ and~\S#2#1#3}
\crefname{example}{Ex.}{Exs.}
\crefname{item}{item}{items}
\crefname{footnote}{footnote}{footnotes}
\crefname{observation}{Obs.}{Obs.}
\crefname{remark}{Remark}{Remarks}
\crefname{proposition}{Prop.}{Props.}
\crefname{equation}{Eqn.}{Eqns.}
\crefname{counterexample}{Counterexample}{Counterexamples}
\crefname{property}{Property}{Properties}
\crefname{algorithm}{Algorithm}{Algorithms}
\crefname{lstlisting}{Listing}{Listings}

\usepackage{shortcuts}

\definecolor{codegreen}{rgb}{0,0.6,0}
\definecolor{codeblue}{rgb}{0,0,0.8}
\definecolor{codegray}{rgb}{0.5,0.5,0.5}
\definecolor{codepurple}{rgb}{0.58,0,0.82}
\definecolor{backcolour}{rgb}{0.95,0.95,0.92}

\lstdefinelanguage{cstar}[]{C}{
    commentstyle=\color{codegreen},
    keywordstyle=\color{codeblue},
    stringstyle=\color{codepurple},
    basicstyle=\ttfamily\footnotesize,
    breakatwhitespace=false,
    breaklines=true,
    captionpos=b,
    keepspaces=true,
    numbers=left,
    numbersep=5pt,
    numberstyle=\tiny,
    showspaces=false,
    showstringspaces=false,
    showtabs=false,
    tabsize=2,
    columns=fullflexible,
    escapechar=@,
    xleftmargin=\leftmargini,
    morekeywords={require, ensure, invariant, assert, parameter, argument, proof, __result, term, thm},
}

\newcommand{\cstar}{\textbf{C$\star$}}

\makeatletter
\renewcommand{\paragraph}{%
  \@startsection{paragraph}{4}%
  {\parindent}{-.2\baselineskip \@plus -2\p@ \@minus -.2\p@}{-3.5\p@}%
  {\bfseries\@parfont\@adddotafter}%
}
\makeatother

\newtheoremstyle{acmremark}%
  {.5\baselineskip\@plus.2\baselineskip\@minus.2\baselineskip}% space above
  {.5\baselineskip\@plus.2\baselineskip\@minus.2\baselineskip}% space below
  {\itshape}% body font
  {\parindent}% indent amount
  {\itshape}% head font
  {.}% punctuation after head
  {.5em}% spacing after head
  {\thmname{#1}\thmnumber{ #2}\thmnote{ {(#3)}}}% head spec
\AtEndPreamble{%
  \theoremstyle{acmplain}%
  \newtheorem*{theorem*}{Theorem}%
  \newtheorem*{lemma*}{Lemma}%
  \newtheorem*{proposition*}{Proposition}%
  \newtheorem*{principle*}{Principle}%
  \theoremstyle{acmdefinition}%
  \newtheorem*{background*}{Background}%
  \theoremstyle{acmremark}%
  \newtheorem{remark}[theorem]{Remark}%
  \theoremstyle{acmplain}%
}

\AtBeginDocument{%
  \setlength\abovedisplayskip{0.5\abovedisplayskip}%
  \setlength\belowdisplayskip{0.5\belowdisplayskip}%
  \setlength\abovedisplayshortskip{0.5\abovedisplayshortskip}%
  \setlength\belowdisplayshortskip{0.5\belowdisplayshortskip}%
  \setlength\floatsep{0.5\floatsep}%
  \setlength\abovecaptionskip{0.5\abovecaptionskip}%
}
\setlist{leftmargin=2\parindent}

\begin{document}

\title{\texorpdfstring{\cstar{}}{C*}: Unifying Programming and Verification in C}
%TODO: CStar: embedding proof code in C? (proposed by Xiong)
%\subtitle{Bringing systems programmers into verification practices}

\author{Yiyuan Cao}
\affiliation{%
  \institution{Peking University}
  \city{Beijing}
  \country{China}
}

\author{Jiayi Zhuang}
\affiliation{%
  \institution{Peking University}
  \city{Beijing}
  \country{China}
}

\author{Houjin Chen}
\affiliation{%
  \institution{Peking University}
  \city{Beijing}
  \country{China}
}

\author{Jinkai Fan}
\affiliation{%
  \institution{Peking University}
  \city{Beijing}
  \country{China}
}

\author{Wenbo Xu}
\affiliation{%
  \institution{Peking University}
  \city{Beijing}
  \country{China}
}

\author{Zhiyi Wang}
\affiliation{%
  \institution{Peking University}
  \city{Beijing}
  \country{China}
}

\author{Di Wang}
\affiliation{%
  \institution{Peking University}
  \city{Beijing}
  \country{China}
}

\author{Qinxiang Cao}
\affiliation{%
  \institution{Shanghai Jiao Tong University}
  \city{Shanghai}
  \country{China}
}

\author{Yingfei Xiong}
\affiliation{%
  \institution{Peking University}
  \city{Beijing}
  \country{China}
}

\author{Haiyan Zhao}
\affiliation{%
  \institution{Peking University}
  \city{Beijing}
  \country{China}
}

\author{Zhenjiang Hu}
\affiliation{%
  \institution{Peking University}
  \city{Beijing}
  \country{China}
}

\renewcommand{\shortauthors}{Trovato et al.}

\begin{abstract}
Ensuring the correct functionality of systems software, given its safety-critical and low-level nature, is a primary focus in formal verification research and applications.
Despite advances in verification tooling, conventional programmers are rarely involved in the verification of their own code, resulting in higher development and maintenance costs for verified software.
A key barrier to programmer participation in verification practices is the disconnect of environments and paradigms between programming and verification practices, which limits accessibility and real-time verification.

We introduce \cstar{}, a proof-integrated language design for C programming.
\cstar{} extends C with verification capabilities, powered by a symbolic execution engine and an LCF-style proof kernel.
It enables real-time verification by allowing programmers to embed proof-code blocks alongside implementation code, facilitating interactive updates to the current proof state.
Its expressive and extensible proof support allows users to build reusable libraries of logical definitions, theorems, and programmable proof automation.
Crucially, \cstar{} unifies implementation and proof code development by using C as the common language.

We implemented a prototype of \cstar{} and evaluated it on a representative benchmark of small C programs and a challenging real-world case study: the \lstinline|attach| function of pKVM's buddy allocator.
Our results demonstrate that \cstar{} supports the verification of a broad subset of C programming idioms and effectively handles complex reasoning tasks in real-world scenarios.
\end{abstract}
\begin{CCSXML}
<ccs2012>
 <concept>
  <concept_id>00000000.0000000.0000000</concept_id>
  <concept_desc>Do Not Use This Code, Generate the Correct Terms for Your Paper</concept_desc>
  <concept_significance>500</concept_significance>
 </concept>
 <concept>
  <concept_id>00000000.00000000.00000000</concept_id>
  <concept_desc>Do Not Use This Code, Generate the Correct Terms for Your Paper</concept_desc>
  <concept_significance>300</concept_significance>
 </concept>
 <concept>
  <concept_id>00000000.00000000.00000000</concept_id>
  <concept_desc>Do Not Use This Code, Generate the Correct Terms for Your Paper</concept_desc>
  <concept_significance>100</concept_significance>
 </concept>
 <concept>
  <concept_id>00000000.00000000.00000000</concept_id>
  <concept_desc>Do Not Use This Code, Generate the Correct Terms for Your Paper</concept_desc>
  <concept_significance>100</concept_significance>
 </concept>
</ccs2012>
\end{CCSXML}


% software verification, 
\keywords{software verification, real-time verification, C programming, LCF-style theorem proving, separation logic, symbolic execution}

\received{20 February 2007}
\received[revised]{12 March 2009}
\received[accepted]{5 June 2009}

\maketitle

\section{Introduction}
\label{sec:introduction}

\paragraph{Background} 
Systems software forms the infrastructure of modern computing, providing the low-level foundation on which all higher-level applications operate.
Given its critical role, 
%ensuring the functional correctness of systems software is paramount.
%
%To achieve a high level of confidence in functional correctness, 
recent years have seen considerable advances in the formal verification of systems software components~\cite{DBLP:conf/fm/LeinenbachS09, DBLP:conf/sosp/TaoYLLNG21, DBLP:conf/sp/LiLGNH21, DBLP:journals/tocs/KleinAEMSKH14, DBLP:conf/cav/XuFFZZL16,DBLP:conf/osdi/GuSCWKSC16, DBLP:conf/asplos/AmaniHCRCOBNLST16, DBLP:conf/sosp/ChenZCCKZ15, DBLP:journals/cacm/Leroy09, DBLP:conf/popl/KumarMNO14, DBLP:conf/sp/ProtzenkoPFHPBB20, DBLP:conf/uss/RamananandroDFS19}.

In this paper, we focus on the verification of software implemented in the C programming language, which remains widely used due to its predictable performance, fine-grained control over system resources, and the vast amount of existing critical code written in it.
Significant progress has been made in verification frameworks and toolchains for C programs~\cite{DBLP:conf/pldi/GreenawayLAK14, DBLP:journals/pacmpl/ZhouQWAC24, DBLP:journals/pacmpl/ManskyD24, DBLP:journals/cacm/Leroy09, DBLP:conf/esop/Appel11, DBLP:conf/pldi/SammlerLKMD021, DBLP:journals/pacmpl/PulteMSMSK23, DBLP:conf/nfm/JacobsSPVPP11,DBLP:journals/pacmpl/GruetterFC24, DBLP:journals/fac/KirchnerKPSY15, DBLP:conf/tphol/CohenDHLMSST09, DBLP:journals/pacmpl/ProtzenkoZRRWBD17}.
The substantial progress in the development of formally verified software components and verification tools has demonstrated the feasibility of large-scale verification, and has brought us closer to the vision where all critical software should be verified~\cite{DBLP:journals/csur/HoareMLS09}.

Despite these successes, formally verified software projects remain costly, in the sense that they require specialized teams with significant expertise, and usually need person-years to complete~\cite{DBLP:journals/cacm/Leroy09, DBLP:journals/tocs/KleinAEMSKH14}.
For the wider adoption of verification practices, the development and maintenance costs for verified software must be reduced.
One significant source of the high costs arises from the \emph{lack of involvement from programmers}, who carry out most of the implementation work yet rarely participate in the verification of their own code.

%The high costs arise in two-folds: (i) the formal reasoning itself requires substantial effort to develop specifications and proofs, and (ii) the non-unified tooling for programming and verification incurs significant overheads in the collaboration between systems programmers and verification experts.

\paragraph{Existing approaches}
One reason for the lack of programmer involvement is that the verification of C programs typically requires an external environment, e.g., an interactive theorem prover such as Coq, which demands programmers to learn a significantly different proving paradigm from the C programming experience.
Examples in this category include AutoCorres~\cite{DBLP:conf/itp/GreenawayAK12}, VST~\cite{DBLP:conf/esop/Appel11} and its recent variant in Iris~\cite{DBLP:journals/pacmpl/ManskyD24}, as well as the Live Verification framework~\cite{DBLP:journals/pacmpl/GruetterFC24}.
The first three translate existing C programs into certain logical representations in the meta-logic (a monadic shallow embedding or a deep embedding) and then require programmers to conduct proofs around these representations in their underlying theorem prover.
The Live Verification framework chooses another approach by composing the program lazily and incrementally along the proof process, relying on the specific mechanism of existential meta-variables in Coq to represent a partially constructed program.

To encourage more involvement from programmers, other C verification tools provide language-level integration of programming and verification to make the verification process more accessible to programmers.
There are two main categories: (i) \emph{assertion-based verifiers} such as Frama-C~\cite{DBLP:journals/fac/KirchnerKPSY15} and VST-A~\cite{DBLP:journals/pacmpl/ZhouQWAC24}, and (ii) \emph{advanced-type-based verifiers} such as RefinedC~\cite{DBLP:conf/pldi/SammlerLKMD021} and CN~\cite{DBLP:journals/pacmpl/PulteMSMSK23}.
These tools allow programmers to annotate a C program with intermediate assertions (e.g., loop invariants) or advanced types (e.g., ownership and refinement types) apart from the specifications to guide the verification process.
These tools are usually \emph{(semi-)automated}, in the sense that they employ an assertion or type checker to automatically verify if the program conforms to the specifications with the help of the programmer-provided annotations.
However, when automation falls short, programmers again need to switch to an external theorem proving environment to complete the verification (e.g., in Coq~\cite{DBLP:journals/pacmpl/ZhouQWAC24, DBLP:conf/pldi/SammlerLKMD021, DBLP:journals/pacmpl/PulteMSMSK23}).
To mitigate the issue, VeriFast~\cite{DBLP:conf/nfm/JacobsSPVPP11}---an assertion-based automated verifier for C programs---provides limited proof support such that programmers can annotate the program with a fixed set of proof commands and write ghost lemma functions to perform certain forms of inductive reasoning.
However, VeriFast lacks the expressiveness and extensibility in proof support required for the collaborative verification of low-level systems software between programmers and proof experts.

\paragraph{Our goal}
As discussed above, there is no satisfactory verification tooling for conventional systems programmers.
In this paper, we aim to design and implement a new C verification tool that satisfies the following two criteria:
\begin{itemize}
  \item it should provide \emph{language-level integration} of programming and verification; and
  \item it should provide \emph{comprehensive proving capabilities} within C's programming paradigm.
\end{itemize}
To further enhance the usability of the C verification tool, we consider one more criterion:
\begin{itemize}
  \item it should provide support for \emph{real-time verification}, i.e., the tool should be able to provide a static summary of the program state at every program point and allow programmers to inspect every intermediate proof state inside a proof.
\end{itemize}

\paragraph{Our approach}
In this paper, we propose \cstar{}, a proof-integrated language that embeds full-fledged verification and proving capabilities in C.
We highlight the three key designs of \cstar{} below.
\begin{itemize}
  \item We adapt the assertion-based design by allowing programmers to annotate a program with \emph{separation-logic} assertions and incorporating \emph{forward symbolic execution}, which abstracts the complexities of concrete semantics and maintains a static summary of the symbolic program state after processing a program fragment.
  \item We integrate the C programming language with \emph{LCF-style proof support} for \emph{higher-order logic}, which provides a comprehensive and extensible interface for programming formal proofs and transforming symbolic states, facilitating the development of high-level reasoning abstractions---as \emph{proof support libraries}---using the full power of C.
  \item With the previous two designs, \cstar{} is ready to support real-time verification: forward symbolic execution provides a summary of the symbolic program state at every program point, and the LCF-style proof support allows programmers to inspect and manipulate the proof state using the familiar programming constructs of C. 
  % TODO: also the ability to transform the symbolic state.
\end{itemize}

%We achieve this by integrating the C programming language with (i) \emph{forward symbolic execution} for unifying the timing, and (ii) \emph{LCF-style proof support} for unifying the paradigm, two well-established components used in program verification and theorem proving:
%%
%\begin{itemize}
%    \item The former abstracts the complexities of concrete semantics based on a separation logic for C, maintaining a static summary of the symbolic program state after processing a program fragment, enabling predictable and operational reasoning in real-time.
%    %
%    \item The latter provides an extensible interface for programming formal proofs in the higher-order logic, facilitating the development of high-level reasoning abstractions using the full power of the proof language, which in our case is also C.
%\end{itemize}
%\cyy{rephrase evaluation results.}
We implemented a prototype of \cstar{} and evaluated it on a suite of C programs to demonstrate \cstar{}'s practicality for the development of verified programs.
Specifically, our evaluation shows that \cstar{} (i) supports systems programming idioms and a large subset of C language features, (ii) provides sufficient expressiveness for advanced ownership and functional reasoning, and (iii) is capable of verifying realistic C programs.
% TODO: weaken the description of the evaluation.
\paragraph{Contributions}
In this paper, we make the following contributions:
\begin{itemize}
    \item We propose a proof-integrated language design that embeds specifications and proof code in C programs, provides comprehensive reasoning capabilities within C's programming paradigm, and thus make formal verification practices accessible to programmers.
    \item We implemented our design as the \cstar{} toolchain by extending C with two well-established components: a symbolic-execution engine and an LCF-style proof kernel, interfacing both to create a lightweight yet powerful verification workflow.
    \item We evaluated our implementation of \cstar{} on a suite of benchmark programs from the literature and a realistic case study to show it is effective in developing verified C programs with the help of \cstar{}'s standard proof-support library.
\end{itemize}

%\paragraph{Roadmap}
%%
%The remainder of this paper is organized as follows.
%%
%In \cref{sec:background}, we provide enough background on the two core components that together make unifying programming and verification of C programs within \cstar{} possible.
%%
%In \cref{sec:overview}, we offer a guided tour of how programmers develop verified C programs in \cstar{}, highlighting its natural development workflow and accessibility to conventional programmers.
%%
%In \cref{sec:design}, we detail the core design of \cstar{}, explaining its (i) syntax and proving capabilities, and (ii) compilation and execution process of proofs, including the interaction between the verification components within the proof environment.
%%
%Next, in \cref{sec:evaluation}, we evaluate \cstar{}, highlighting its extensibility in reasoning capabilities and practicality in verified systems software development.
%%
%In \cref{sec:relatedwork}, we discuss related work in the field.
%%
%Finally, in \cref{sec:conclusion}, we conclude the paper with a summary, along with limitations and directions for future work.
%\input{background}
\section{A Guided Tour of \texorpdfstring{\cstar{}}{C*}}
\label{sec:overview}

%\begin{figure}[t]
%\centering
%\includegraphics[width=0.75\textwidth]{cstar-concept.pdf}
%\caption{A conceptual illustration of the collaboration between programmers and proof experts in \cstar{}.}
%\label{fig:cstar-concept}
%\end{figure}

% In CStar, programmers and proof experts collaborate to develop verified software.
% This section shows the programmer's view of the verified program development.
% They write C code together with proof code that uses derived rules and operational reasoning abstractions (not shown in this example) built by proof experts.
% We argue that the programmer's reasoning can be conveyed at high-level in this way.

%Using \cstar{}, programmers and proof experts can collaborate in the development of verified software in a single environment as illustrated in \cref{fig:cstar-concept}.
%%
%Programmers write implementation code and proof code in the same language: when a piece of implementation code whose behavior needs non-trivial reasoning is written, they can add \emph{proof-code blocks} to reason about and transform the symbolic state in real-time, using the programming constructs of C, supported by the high-level proof functions and more built by proof experts.
%%
%Proof experts help by developing reusable \emph{proof-support libraries}, containing logical definitions, theorems, and programmable proof automation, providing high-level reasoning capabilities accessible to the programmers, possibly tailored to specific domains at hand.

In this section, we present a guided tour for developing a verified C program in \cstar{}.
We take the \lstinline|clear| function shown in \cref{lst:clear_snapshot} as the running example, whose desired functionality is to reset \lstinline|len| contiguous bytes that start from a base address \lstinline|to|.
The implementation code of \lstinline|clear| consists of lines 3, 6, 8, 10, 17, 19, 20, 22, and 24; others are verification-specific code.
In the implementation code, the programmer declares a local variable \lstinline|i| in line 8, followed by a loop from line 10 to line 22, where each loop iteration sets the \lstinline|i|-th byte from the base address \lstinline|to| to zero and increments \lstinline|i| by one until \lstinline|i| reaches the function parameter \lstinline|len|.

We explain the verification-specific code in a way that guides the reader through the incremental development process following \cstar{}'s workflow.
Our explanation will follow the three criteria mentioned in \cref{sec:introduction}:
\cref{sec:overview:integration} for language-level integration of programming and verification, e.g., how the user writes specifications and assertions about \lstinline|clear|;
\cref{sec:overview:proving} for comprehensive proving capabilities via C programming,
e.g., how the user proves the implementation of \lstinline|clear| comforms to its specification;
and \cref{sec:overview:realtime} for real-time program verification, e.g., how \cstar{} aids the user during the incremental development.

\begin{figure}[t]
\begin{lstlisting}[
    caption={An example verified C program in \cstar{}.},
    label={lst:clear_snapshot},
]
#include "cstarlib.h"
#include "clear.h"
void clear(void *to, int len)
    [[require(`fact(len >= 0) ** undef_array_at(to, Tchar, len)`)]]
    [[ensure(`array_at(to, Tchar, replicate(len, 0))`)]]
{
    @«@ term params = `data_at(&"to", Tptr, to) ** data_at(&"len", Tint, len)`; @»@
    int i = 0;
    @«@ /* proof: establish invariant */ @»@
    while (i < len)
        [[invariant(`@$\exists$@(i:integer).
            fact(0 <= i && i <= len) **
            data_at(&"i", Tint, i) ** ${params:hprop} **
            array_at(to, Tchar, replicate(i, 0)) ** 
            undef_array_at(to + i * sizeof(Tchar), Tchar, len - i)
        `)]]
    {
        @«@ single_out_location(); @»@
        *((char *)to + i) = (char) 0;
        i = i + 1;
        @«@ /* proof: re-establish invariant */ @»@
    }
    @«@ /* proof: establish post-condition */ @»@
}
\end{lstlisting}
\end{figure}
% TODO: get_symbolic_state() and set_symbolic_state() are context-sensitive. Could it be a problem here?

\subsection{Language-level Integration of Programming and Verification}
\label{sec:overview:integration}

The first thing to do for verifying a program is correct is to \emph{specify} how it is supposed to be correct.
It is a common practice to specify a function's expected behavior by formulating its \emph{pre-} and \emph{post-}conditions, i.e., the expected program states before calling the function and after returning from it.
For the desired functionality of \lstinline|clear|, the pre-condition could be that the function parameter \lstinline|len| is non-negative and the other parameter \lstinline|to| is a base address that points to a memory chunk of at least \lstinline|len| contiguous bytes.
The corresponding post-condition would be that \lstinline|len| contiguous bytes starting from the address \lstinline|to| are set to zero.
To formally formulate such conditions for low-level heap-manipulating programs such as \lstinline|clear|, we adapt \emph{separation logic} in \cstar{}'s design.

\begin{background*}[Separation Logic]
The development of separation logic~\cite{DBLP:journals/cacm/OHearn19, DBLP:conf/lics/Reynolds02} is driven by the desire to verify heap-manipulating low-level programs in a modular manner, especially for handling the flexibility of aliasing.
Its most salient features are (i) the introduction of a logical connective, \emph{separating conjunction}, expressing non-aliasing properties between heap fragments in a succinct way, and (ii) a characterizing program proof rule, \emph{frame rule}, extending the \emph{program locality} of Hoare logic rules with \emph{spatial locality} when reasoning about heap-manipulating programs.
\end{background*}

\cref{fig:assertion-syntax} lists the concrete notations for some separation-logic predicates used in \cstar{}.
%
%Standard logic predicates are lifted to separation-logic predicates with notations overloaded (e.g., \lstinline[language=C]|&&| for conjunction and \lstinline[language=C]$||$ for disjunction).
%
The standard proof-support library of \cstar{} provides some other widely-used predicates for program verification tasks.
For example,
\lstinline|array_at(p, ty, lst)| represents a consecutive array of elements of C type \lstinline|ty| starting at the address \lstinline|p|, where the $n$-th element of the array is represented by the $n$-th element in the logic-level list \lstinline|lst|.
Another example 
 \lstinline[language=C]|undef_array_at(p, ty, len)| represents an array starting at address \lstinline[language=C]|p| with \lstinline[language=C]|len| undefined values, each of which is uninitialized or irrelevant to the verification.
With separation logic, the user can specify the pre- and post-conditions in lines 4 and 5, respectively:
\begin{itemize}
  \item \cstar{} uses the C attribute syntax \lstinline|[[require]]|\footnote{In actual code, all the attributes are prefixed with the \lstinline|[[cstar::]]| namespace for disambiguation.} to enclose a pre-condition. In line 4, the predicate \lstinline|fact(len >= 0)| represents an \emph{empty} heap with the condition that \lstinline|len| is non-negative. The predicate \lstinline|undef_array_at(to, Tchar, len)| represents a memory chunk of \lstinline|len| continuous bytes, where \lstinline|Tchar| is the logic-level representation of the C type \lstinline|char|.
      Using separating conjunction \lstinline|**| to compose the two predicates yields a precise formulation of the intended pre-condition.
  \item \cstar{} uses \lstinline|[[ensure]]| for post-conditions.
      In line 5, the predicate \lstinline|array_at(to, Tchar, replicate(len, 0))| represents a memory chunk of \lstinline|len| continuous zeros, where
      the logic-level term \lstinline|replicate(len, 0)| creates a list of \lstinline|len| zeros.
      This, again, precisely corresponds to the intended post-condition we discussed earlier.
\end{itemize}

\begin{table}[t]
    \centering\small
    \caption{Concrete notations for some separation-logic predicates used in \cstar{}.}
    \label{fig:assertion-syntax}
    \begin{threeparttable}
    \begin{tabular}{|l|l|l|}
        \hline
        \textbf{Separation-logic Predicate} & \textbf{Concrete Notation} & \textbf{(Simplified) Definition}
        %\tnote{1} 
        \\
        \hline
        empty predicate & \lstinline[language=C]|emp| & $\lambda h.\ h = \emptyset$ \\[4pt]
        \multirow{2}{*}{embedded proposition} & \lstinline[language=C]|fact(p)| & $\lambda h.\ h = \emptyset \land p$ \\
        & \lstinline[language=C]|pure(p)| & $\lambda h.\ p$ \\[4pt]
        \multirow{2}{*}{singleton maps-to} & \lstinline[language=C]|data_at(x, ty, v)| & $\lambda h.\ h = (x \mapsto_{\textrm{ty}} v) \land valid_{\textrm{ty}}(x, v)$ \\
        & \lstinline[language=C]|undef_data_at(x, ty)| & $\lambda h.\ h = (x \mapsto_{\textrm{ty}} \_) \land valid_{\textrm{ty}}(x, \_)$ \\[4pt]
        separating conjunction & \lstinline[language=C]|hp1 ** hp2| & $\lambda h.\ \exists h_1\, h_2.\, h_1 \uplus h_2 = h \land hp_1\, h_1 \land hp_2\, h_2$ \\
%        existential quantifier & \lstinline[language=C]|exists x. hp| & $\lambda h.\ \exists x.\, hp\, h$ \\
        \hline
    \end{tabular}
    % \begin{tablenotes}
    %     \item[1] Currently the predicates are axiomatized in \cstar{}, and will be translated to corresponding assertion syntax used by the external symbolic execution engine when the program is being symbolically executed.
    % \end{tablenotes}
    \end{threeparttable}
\end{table}

Readers may have noticed the uses of \emph{quotations} \lstinline|`...`| inside pre- and post-conditions.
The quotation mechanism allows the user to construct separation-logic predicates and other logic-level terms using conventional concrete syntax, which is similar to existing assertion-based C verifiers.
But as we will show in \cref{sec:overview:proving}, these terms are \emph{first-class} values in \cstar{}:
beyond being directly written with quotations, they can be computed from expressions, stored in variables, passed as arguments, and manipulated using the full capabilities of the C programming language.
This is one key difference between \cstar{} and traditional assertion-based C verifiers.

Writing pre- and post-conditions is far from completing the verification, because it is generally intractable to have an algorithm to automatically verify the function body ``transforms'' the pre-condition to the post-condition.
Similar to many existing C verifiers, \cstar{} adapts the assertion-based design to enable
a \emph{declarative} style of verification:
\begin{principle*}[Declarative Style of Verification]
The user annotates the program with separation-logic assertions about the expected program states that hold at specific program points.
\end{principle*}
\noindent A particular important class of assertions are \emph{loop invariants}.
\cstar{} also uses the C attribute \lstinline|[[invariant]]| to accompany a loop with its invariant, i.e., a separation-logic predicate that is expected to hold at the beginning of each loop iteration.
Lines 11--16 specify the loop invariant, which intuitively states that at the \lstinline|i|-th iteration,
the value of \lstinline|i| should be between zero and \lstinline|len| (line 12),
local variables and parameters are stored properly in the memory with \lstinline|Tint| being the logic-level representation of the C type \lstinline|int| (line 13), and
the base address \lstinline|to| points to a memory chunk that
starts with \lstinline|i| zeros (line 14) followed by \lstinline|len - i| unspecified bytes (line 15).
Note that in line 13 the code uses an \emph{anti-quotation} \lstinline|${params:hprop}| to interpolate a predicate defined in line 7.
This again indicates that predicates are first-class values and we defer the discussion of anti-quotations to \cref{sec:overview:proving}.

With extra assertions including invariants, an assertion-based verifier usually splits the verification into multiple sub-tasks, each of which corresponds to prove a Hoare triple for a \emph{straight-line} program segment.
For example, to verify that the \lstinline|clear|'s implementation code conforms to the pre- and post-conditions, it is sufficient to complete three sub-tasks (i.e., verifying three Hoare triples):
% Note: applying the consequence rule can be seen as proving triples for the idle command (;).
\begin{itemize}
  \item prove the code in line 8 transforms the pre-condition to the loop invariant (line 9);
  \item prove that the loop body (lines 19 and 20) re-establishes the loop invariant (line 21); and
  \item prove that the loop invariant---with the loop condition (in line 10) being false---entails the post-condition (line 23).
\end{itemize}
Each sub-task, i.e., the proof of each Hoare triple $\{~P~\}~S~\{~Q~\}$, involves two parts: (i) reasoning about semantics of the program $S$, i.e., finding the strongest post-condition $Q_\mathsf{sp}$ of $S$ w.r.t. the pre-condition $P$,
and (ii) carrying out an entailment proof, i.e., proving that $Q_\mathsf{sp}$ entails $Q$.
Instead of employing automated provers for both parts---as many other assertion-based verifiers do---\cstar{} adapts a \emph{predictable} mechanism of automation by integrating \emph{forward symbolic execution} to reason about program semantics, i.e., part (i) of each verification sub-task.

\begin{background*}[Forward Symbolic Execution]
Since the work of Berdine et al.~\cite{DBLP:conf/aplas/BerdineCO05}, separation logic has been effectively used as a sound foundation for forward symbolic execution, which closely matches a programmer's operational intuition about the effects of statements on program states, while abstracting away concrete semantic details.
It achieves this by providing a highly predictable algorithm for automatically applying structural program logic rules for separation-logic predicates in suitable forms, i.e., the \emph{symbolic heap} fragment of separation logic~\cite{DBLP:journals/jar/CaoBGDA18}.
\end{background*}

Using a symbolic-execution engine, \cstar{} computes a \emph{symbolic state} for each program point.
The symbolic state consists of the values of the program variables and the view of the heap fragments that are worked on and owned by the program.
% , also known as its \emph{memory footprint}.
%
For example, at the beginning of the function body of \lstinline|clear| (in line 6), symbolic execution uses the pre-condition annotated in line 4 to initialize the symbolic state to be the same as the following separation-logic predicate, which additionally consists of \lstinline|data_at| predicates for the function parameters:
%\footnote{In actual code, the initial value of the parameters are referred to using the \lstinline|@pre| syntax, such as \lstinline|to@pre||. For simplicity, we omit the \lstinline|@pre| syntax in the snapshot.}
%In actual code, all scalar C values are represented using the integer type (Z or int). And literals in HOL Light are by default of type :num, so we need to convert them to the correct type using the \lstinline|int_of_num| function, and &0 is the notation for this conversion.
\begin{lstlisting}[numbers=none]
fact(len >= 0) ** undef_array_at(to, Tchar, len) **
  data_at(&"to", Tptr, to) ** data_at(&"len", Tint, len)
\end{lstlisting}
Here \lstinline|Tptr| is the logic-level representation of C's pointer types.

If symbolic execution were always successful, the remaining proof obligations for the user would all be separation-logic entailments, i.e., part (ii) of each verification sub-task.
In \cref{sec:overview:proving}, we will show \cstar{}'s capabilties in supporting the user to develop the logical proofs.
On the other hand, unfortunately, the price of having a predictable symbolic-execution engine is that it will not try to automate the reasoning and transformations on the symbolic state that a user might find intuitive to perform.
For example, in line 19, the statement assigns to the address \lstinline|((char *)to + i)|, but the symbolic state---the loop invariant in this case---does not explicitly describe the memory cell pointed by the address.
As a result, \cstar{}'s symbolic-execution engine cannot (yet) automatically process the assignment.
In \cref{sec:overview:proving}, nevertheless, we will show how \cstar{}'s proving capabilities provide an \emph{operational} style of verification, where users can convey and formalize their high-level intuitive ideas on manipulating the symbolic state.

\subsection{Comprehensive Proving Capabilities via C Programming}
\label{sec:overview:proving}

As discussed in \cref{sec:overview:integration}, \cstar{}'s proving capabilities should support its users in the following two tasks:
\begin{itemize}
  \item developing logical proofs for entailments, and
  \item manipulating symbolic states inside the implementation code.
\end{itemize}
In particular, \cstar{}'s support should satisfy a key criterion:
\begin{itemize}
  \item allow the user to \emph{programmably} develop logical proofs and manipulate symbolic states, using C's conventional programming constructs.
\end{itemize}
To achieve the aforementioned goals, we adapt \emph{LCF-style theorem proving} in \cstar{}.

\begin{background*}[LCF-style Theorem Proving]
The LCF architecture is a general technique for embedding formal logics into a programming language.
Pioneered by Robin Milner and colleagues in the early work on the Edinburgh LCF theorem prover~\cite{DBLP:conf/birthday/Gordon00, DBLP:series/hhl/HarrisonUW14}, its descendants are still widely used today~\cite{DBLP:conf/tphol/Harrison09a, DBLP:books/sp/NipkowPW02}.
In LCF-style provers, a general-purpose programming language is used as the \emph{meta-language} to implement \emph{object logic} entities such as \emph{terms}, \emph{types}, and \emph{theorems}.
These are represented as recursive data structures, making formal proof a \emph{programming} process of constructing theorems from a set of axioms using primitive inference rules.
Specifically, the axioms are encoded as constants, and the inference rules are implemented as functions that take premises and return conclusions as theorems if the rules can be successfully instantiated.
\end{background*}

In \cstar{}, we integrate an LCF-style proof kernel with \emph{higher-order logic} as the object logic.
The kernel is wrapped by a C interface; in other words, the meta-language in \cstar{}'s design is the standard C programming language.
To distinguish the code for developing logical proofs and manipulating symbolic states from ordinary implementation code, 
\cstar{} introduces \emph{proof-code blocks} delimited by the «\lstinline|...|» syntax.\footnote{In actual code, we use the \lstinline|[[cstar::proof(...)]]| attribute to embed proof-code blocks.}
Arbitrary C code is allowed in proof-code blocks, with the ability to introduce bindings and construct values of type \lstinline|term| and \lstinline|thm|, corresponding to object-logic terms and theorems, respectively.
Recall that we mentioned that in \cstar{}, separation-logic predicates are first-class values.
Indeed, they are just \lstinline|term| values of object-logic type \lstinline|hprop| (short for heap propositions).
The quotation and anti-quotation mechanisms are thereby introduced to conveniently construct \lstinline|term| values.
For example, in line 7 of \cref{lst:clear_snapshot}, the code stores the \lstinline|data_at| predicates regarding ownership of the parameters---wrapped by a quotation \lstinline|`...`|---in a variable called \lstinline|params|.
In the following proof-code blocks and assertions (e.g., invariants), the user can use \lstinline|params| as if it is a normal program variable.\footnote{All proof blocks in a function body are in the same scope, and global proof-code blocks (that is outside of any function body) are file-scoped. Local scopes can be created using C blocks \lstinline|\{...\}|. See \cref{sec:implementation} for more details.}
In line 13, the code indeed uses it;
combined with the anti-quotation \lstinline|${params:hprop}|, it reduces redundancy when writing the loop invariant.

Because separation-logic predicates are just values in \cstar{}, it becomes natural to write C code to manipulate symbolic states, which are special kinds of separation-logic predicates.
Such capability of \cstar{} enables an \emph{operational} style of verification:
\begin{principle*}[Operational Style of Verification]
  The user manipulates the symbolic state using arbitrary C code, provided they can give justifications, i.e., theorems for the corresponding separation-logic entailments, for the manipulations they made.
\end{principle*}
\noindent For example, the operational style of verification is applied in line 18.
Here, the symbolic state---computed by the symbolic-execution engine---is equivalent to the following predicate:
%\footnote{
%% There are minor syntactic differences in the assertion format written in \cstar{} and the format used by the symbolic executor, but converting between the two notations is straightforward.
%%
%Not identical because anti-quotation is not supported by the symbolic executor, so they will always be expanded in full in the symbolic state returned by the symbolic executor. See \cref{sec:design} for details about the interaction.}
%
% \begin{figure}[H]
\begin{lstlisting}[numbers=none,xleftmargin=0pt]
@$\exists$@(i:integer). fact(i < len) ** fact(0 <= i && i <= len) **
  data_at(&"i", Tint, i) ** data_at(&"to", Tptr, to) ** data_at(&"len", Tint, len) **
  array_at(to, Tchar, replicate(i, 0)) ** undef_array_at(to + i * sizeof(Tchar), Tchar, len - i)
\end{lstlisting}
% \end{figure}
%
As discussed above about symbolic execution, symbolically executing the next statement (in line 19) would fail, because the symbolic state does not explicitly describe the address \lstinline|((char *)to + i)|.
Intuitively, the user should ``transform'' the symbolic state to some form containing
\lstinline|undef_data_at(to + i * sizeof(Tchar), Tchar)| as a separating conjunct, which represents the ownership of the memory location being stored into.
By inspecting the symbolic state, the user can see that a transformation of the view of the heap is needed to satisfy the requirement: by splitting the \lstinline|undef_array_at| predicate into the separating conjunction of its head element (as an \lstinline|undef_data_at| predicate) and the rest of the slice (as an \lstinline|undef_array_at| predicate with smaller length and starting at a bigger offset).
The split is valid due to the fact that \lstinline|i < len| holds in current state, which entails the fact \lstinline|len - i > 0|, meaning the slice is non-empty.
The corresponding justification---as proof code---for this intuitive transformation is wrapped in the proof procedure \lstinline|single_out_location|.
%
%It demonstrates how \cstar{} allows manipulating the symbolic state using conventional C programming constructs and high-level derived rules from the proof-support libraries, making the intuitive reasoning process easy to implement as proof code.

The implementation of the proof procedure \lstinline|single_out_location| is shown in \cref{lst:single_out_location}.
It demonstrates how \cstar{} allows proving separation-logic entailments using conventional C programming constructs and high-level derived rules from the proof-support libraries, making the intuitive reasoning process easy to implement as proof code.
Thanks to the LCF-style design, logical rules are just C functions that return \lstinline|thm| values, which represent proven theorems.
\begin{itemize}
    \item \emph{Derive the transformation rule (lines 2--7).}
    The \lstinline|undef_array_at_select_first| theorem, from the proof-support library \lstinline|clear.h| included by the \cstar{} program in \cref{lst:clear_snapshot}, asserts that for any uninitialized array, we can single out the first element and treat the remainder as another uninitialized array, provided the length of the array is greater than zero.
    % It specializes the undef_array_at_split theorem.
    The resulting theorem from this derived-rule application is shown by the comment in lines 4--7, where \lstinline|==>| denotes standard logical implication and \lstinline!|--! denotes separation-logic entailment.
    \item \emph{Rewrite using linear arithmetic facts (lines 8--14).}
    The \lstinline|rewrite_rule_list| function---from the standard proof-support library \lstinline|cstarlib.h|---takes an \lstinline|NULL|-terminated array of equational theorems and another theorem, and rewrites the second argument using the equational theorems.
    It is used here to rewrite the theorem using linear arithmetic facts, which are derived automatically by calling the \lstinline|arith_rule| function, to align with the predicates in the current symbolic state.
    The array-destructing theorem \lstinline|undef_array_at_destruct| is re-assigned to the rewritten theorem by the statement in line 13.
    \item \emph{Perform local transformation with frame inferred from the symbolic state (lines 15--17).}
    The \lstinline|local_apply| function is an important derived rule in \lstinline|cstarlib.h|.
    %
    % TODO: cite similar rules in interactive frameworks such as VST (sep_apply), CFML, etc.
    It allows programmers to perform a local transformation with the frame being inferred from the symbolic state.
    Specifically, it takes two arguments: the current symbolic state and a local-transformation theorem.
    %
%    It requires all the separating conjuncts in the antecedent of the entailment to be present in the first assertion, and returns an separation logic entailment theorem that justifies transforming the assertion to the updated assertion where the local transformation justified by the entailment theorem is performed with the frame being inferred from the rest of the assertion.
    %
    In line 15, we fetch the current symbolic state using the built-in \lstinline|get_symbolic_state| function, and use the \lstinline|local_apply| function to perform a local transformation justified by the theorem \lstinline|dest_undef_array|.
    The result of the transformation is then put back to the symbolic-execution engine using the built-in \lstinline|set_symbolic_state| function in line 17.
%    which is a built-in function provided by the \cstar{} compiler that checks the passed-in theorem is indeed a separation-logic entailment from the current symbolic state to another symbolic heap assertion, and if so, informs the symbolic-execution engine to update the current symbolic state to the new one.
    % See \cref{sec:design} for more details.
\end{itemize}

\begin{figure}[t]
    \centering
\begin{lstlisting}[
    caption={Proof procedure: single out the first element of the uninitialized slice.},
    label={lst:single_out_location}
]
void single_out_location(void) {
    thm dest_undef_array =
        undef_array_at_select_first(`to + i * sizeof(Tchar)`, `Tchar`, `len - i`);
      /* len - i > 0 ==>
         undef_array_at(to + i * sizeof(Tchar), Tchar, len - i) |--
         undef_data_at(to + i * sizeof(Tchar), Tchar) **
         undef_array_at((to + i * sizeof(Tchar)) + sizeof(Tchar), Tchar, len - i - 1) */
    thm arith_facts[] = {
        arith_rule(`len - i > 0 <=> i < len`),
        arith_rule(`len - i - 1 == len - (i + 1)`),
        arith_rule(`(to + i * sizeof(Tchar)) + sizeof(Tchar) ==
                    to + (i + 1) * sizeof(Tchar)`), NULL };
    dest_undef_array = rewrite_rule_list(arith_facts, dest_undef_array);
      /* rewrite using linear arithmetic facts */
    thm final_thm = local_apply(get_symbolic_state(), dest_undef_array);
      /* perform local transformation with frame inferred from the symbolic state */
    set_symbolic_state(final_thm); /* update the symbolic state */
}
\end{lstlisting}
\end{figure}

\subsection{Real-time Program Verification}
\label{sec:overview:realtime}

In the previous two sections, we used the verification of the function \lstinline|clear| to illustrate
(i) how \cstar{} incorporates separation logic and forward symbolic execution to provide language-level integration of programming and verification, as well as
(ii) how \cstar{} integrates LCF-style proof support for higher-order logic to provide comprehensive proving capabilities within C's programming paradigm.
In particular, separation-logic predicates are first-class values and proof-code blocks can manipulate symbolic states and proof states.

We claim that \cstar{} achieves \emph{real-time} program verification, i.e., the user can carry out verification as they program the implementation code incrementally.
It achieves this goal by orchestrating the symbolic-execution engine and the LCF-style proof kernel together, creating a \emph{proof-supporting runtime} that runs proof-code blocks and symbolic execution of program segments in an \emph{interleaving} manner, and provides the symbolic state at every program point in implementation code as well as the proof state in proof-code blocks.

Firstly, \cstar{} is capable of providing the symbolic state at every program point, given that (i) every function has pre- and post-conditions, (ii) every loop has an invariant, and (iii) the required maps-to predicates are present in the symbolic state before executing a primitive statement.
This is achieved by a combination of forward symbolic execution for separation logic and the ability to write proof-code blocks to manipulate the symbolic state.
The symbolic state in the symbolic-execution engine is always represented as a separation-logic assertion in a canonical form, known as \emph{symbolic heaps}~\cite{DBLP:conf/aplas/BerdineCO05}.
When a statement (e.g., an assignment) is symbolically executed, the symbolic-execution engine requires the primitive maps-to predicate (i.e., \lstinline|data_at| or \lstinline|undef_data_at|) for the accessed memory locations be present in the current symbolic heap as a separating conjunct; if so, the engine modifies the symbolic heap locally~\cite{DBLP:conf/nfm/JacobsSPVPP11}.
For example, when executing the store statement \lstinline|*((char *)to + i) = (char) 0| in line 19 of \cref{lst:clear_snapshot}, the symbolic-execution engine confirms that the current symbolic heap contains the assertion \lstinline|`undef_data_at(to + i * sizeof(Tchar), Tchar)`|, representing the ownership of the memory location being stored into, and then substitutes the predicate with \lstinline|`data_at(to + i * sizeof(Tchar), Tchar, 0)`|, reflecting the effect of the store statement.
The other predicates in the symbolic heap are left unchanged, being justified by the frame rule of separation logic.
It is worth noting that for the symbolic-execution engine to achieve this, \cstar{} needs to first execute the proof-code block in line 18 of \cref{lst:clear_snapshot} to transform the symbolic heap accordingly, before symbolically executing the store statement.

Secondly, \cstar{}'s runtime environment for running proof-code blocks is capable of providing the proof state in proof-code blocks.
The proof state records the concrete values of \lstinline|term| and \lstinline|thm| variables declared in proof-code blocks, as well as proof functions and theorems included from proof-support libraries.
For example, when a user is developing the function \lstinline|clear| and writes down line 18 of \cref{lst:clear_snapshot} to call \lstinline|single_out_location|, the environment of the proof-supporting runtime should be able to find the function definition of \lstinline|single_out_location| in \cref{lst:single_out_location}.
This ability comes from the LCF-style theorem proving, where proofs are ordinary programs that manipulate terms and theorems.
Thus, \cstar{} can assemble all the proof-code blocks inside a function and its dependent proof-support functions together as a C program, compile it, and execute it to record the concrete values of variables.

% Thirdly, readers may have noticed that the symbolic-execution engine and the LCF-style proof kernel are \emph{not} independent.
% %
% The symbolic-execution engine needs to invoke the proof kernel to execute proof-code blocks to transform symbolic states from time to time,
% and the execution of proof-code blocks needs to appeal to symbolic execution to retrive the symbolic state at a program location.
% %
% Therefore, our design of \cstar{} tightly couples its symbolic execution together with its LCF-style proof kernel as a \emph{proof-supporting runtime}, which provides comprehensive support for proof checking of \cstar{} programs.
% %
% In the next section, we will present the core design of \cstar{}'s runtime as well as its user interface.
\section{Core Design}
\label{sec:design}

In this section, we present the core design of \cstar{} from two perspectives.
Following the discussion in \cref{sec:overview:realtime},
we explain \cstar{}'s internal mechanisms from the developer's perspective in \cref{sec:workflow}, i.e., \cstar{}'s verification-specific workflow and its proof-supporting runtime.
Following the guided tour in \cref{sec:overview:integration,sec:overview:proving}, we describe \cstar{}'s language features from the user's perspective in \cref{sec:design:interface}, i.e., \cstar{}'s verification-specific interface and its proving capabilities.

%\diw{TODO: update this.}
%%
%This section first presents the verification capabilities of \cstar{} that enable programmers and proof experts to develop verified C software and create reusable proof libraries.
%%
%To highlight the extensibility and programmability of the proof support, we walkthrough the implementation of a useful derived rule \lstinline|which_implies| for doing local transformations on the symbolic state.
%%
%We then describes the underlying architecture and an ideal workflow of the \cstar{} toolchain.

% This subsection serves as a quick manual for new users, including the annotation and separation-logic assertion syntax, the primitive types and functions in the proof kernel interface, the built-in functions for accessing the symbolic state, etc.
% TODO: use small examples in between.  SE engine only takes symbolic heaps.
% TODO: provide a figure/table cheat sheet of code snippets.

\subsection{Workflow of \texorpdfstring{\cstar{}}{C*} Toolchain}
\label{sec:workflow}
% TODO: show input and output file excerpts?
% Three components: compiler, proof kernel, symbolic execution engine. (The last two consists the proof-supporting runtime.)
% Two phases: proof-checking phase (in the proof-supporting runtime) and deployment phase (in the standard C runtime).
% Three stages: translation, operational proof checking, residual proof checking.

\begin{figure}[t]
    \centering
    % \textbf{Add a figure for the conceptual workflow (incremental symbolic execution interleaved with proof code execution.).}
    \includegraphics[width=0.8\textwidth]{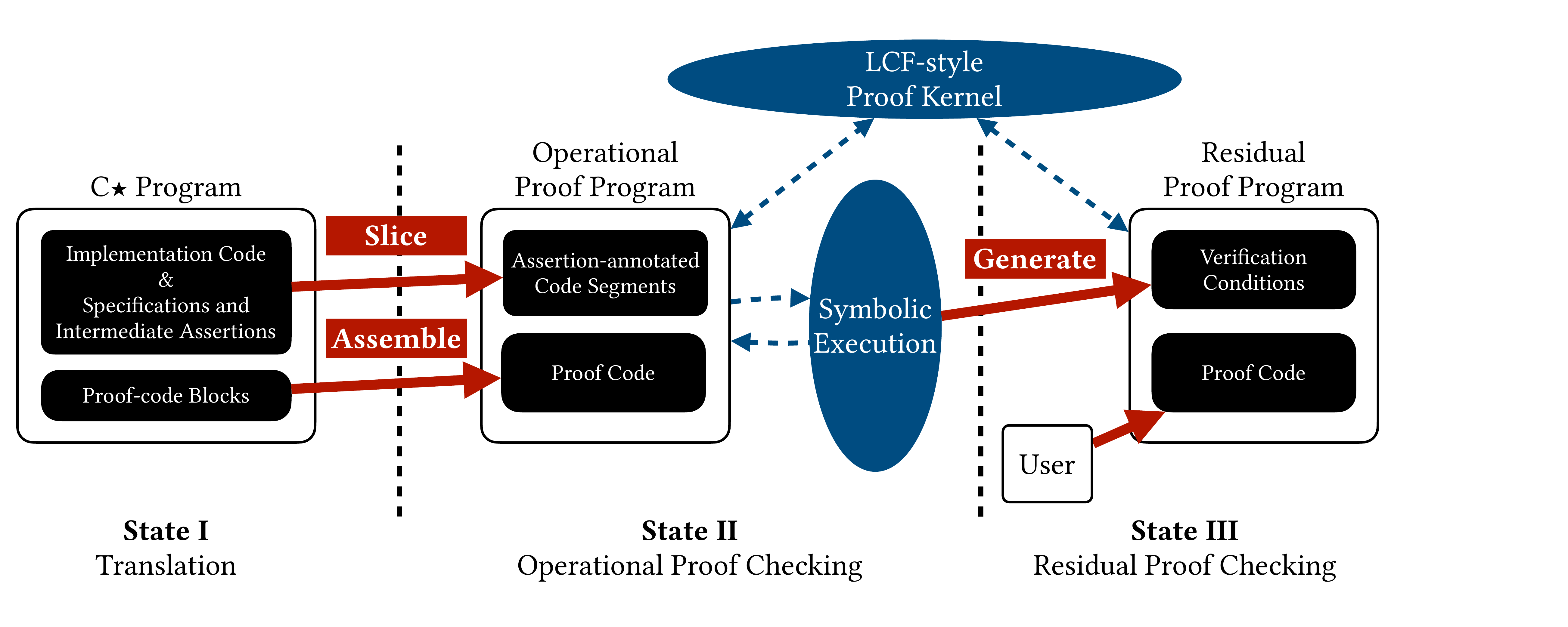}
    \caption{The ideal workflow of \cstar{}'s proof-checking phase.}
    \label{fig:cstar-conceptual-workflow}
\end{figure}

The \cstar{} toolchain consists of three components: the \cstar{} compiler, the LCF-style proof kernel, and the symbolic-execution engine.
The interaction between these components during the \emph{proof-checking phase} of a \cstar{} program is illustrated conceptually in \cref{fig:cstar-conceptual-workflow}.
After proof checking, the \emph{deployment phase} becomes straightforward: because all verification-specific annotations are wrapped in C attributes, the verified program can be compiled directly with C compilers such as \texttt{gcc} or \texttt{clang}.

The proof-checking phase of a \cstar{} program can be summarized as a three-stage process: the \emph{translation stage}, the \emph{operational proof checking stage}, and the \emph{residual proof checking stage}.

\paragraph{Translation stage.}
In the first stage, the \cstar{} compiler processes the input \cstar{} program, and produces an \emph{operational proof program}.
The input \cstar{} program consists of three main components: (i) implementation code, (ii) verification annotations including specifications (e.g., \lstinline|[[require]]| and \lstinline|[[ensure]]| attributes) and intermediate assertions (e.g., \lstinline|[[invariant]]| attributes), and (iii) embedded proof-code blocks.
During translation, the \cstar{} compiler combines part (i) and part (ii) to form the annotated C code, slices it into segments that are separated by proof-code blocks, and stores the segments as serializable data structures in the proof code.
For part (iii), i.e., the proof-code blocks, the \cstar{} compiler assembles them into the main code for execution.
The compiler also handles syntax extensions such as quotation and anti-quotation, translating them into applications of term-constructing functions.
%\footnote{To simplify the description here, we assume that assertions in intermediate assertions do not contain anti-quotations.
% Anti-quotations in intermediate assertions are handled in the implementation using similar treatment of assertions in proof code blocks.
%}
% Small example here?
%
\cref{fig:example-workflow} demonstrates \cstar{}'s workflow using the verification of the \lstinline|clear| function shown in \cref{sec:overview}.
Note that here we reinterpret the commented-out proof-code blocks in lines 9, 21, and 23 of \cref{lst:clear_snapshot} as they are \emph{not} inserted into the implementation code.
\cref{fig:example-operational-proof} is the assembled operational proof program:
lines 1--15 are three code segments split by proof-code blocks in lines 7 and 18 of \cref{lst:clear_snapshot}.
In the main function, we use the built-in function \lstinline|feed_program_segment| to feed a code segment to the symbolic-execution engine, as we will explain below about the second stage.
%
%During translation, the compiler merges the implementation code with the specifications and intermediate assertions to form an \emph{annotated C code}.
%%
%It then slices this code into segments separated by proof code blocks, and stores these segments as string literals in the proof program.
%%
%The proof code blocks themselves are assembled as the main code during the execution of the proof program.
% The assemble process needed here?

\begin{figure}[t]
\centering
\begin{subfigure}[b]{0.54\textwidth}
\begin{lstlisting}[language=C,xleftmargin=0pt]
code_segment_t seg1 = /* 
  void clear(void *to, int len)
    [[require(...)]]
    [[ensure(...)]]
  { */;
code_segment_t seg2 = /*
    int i = 0;
    while (i < len)
      [[invariant(...)]]
    { */;
code_segment_t seg3 = /*
      *((char *)to + i) = (char) 0;
      i = i + 1;
    }
  } */;
int main(void) {
    feed_program_segment(seg1);
    term params = ...; // line 7 of @\color{codegreen}\cref{lst:clear_snapshot}@
    feed_program_segment(seg2);
    single_out_location(); // line 18 of @\color{codegreen}\cref{lst:clear_snapshot}@
    feed_program_segment(seg3);
}
\end{lstlisting}
\caption{Operational Proof Program}\label{fig:example-operational-proof}
\end{subfigure}
\hfil
\begin{subfigure}[b]{0.45\textwidth}
\begin{lstlisting}[language=C,xleftmargin=0pt]
term vc1 = /* establish invariant 
  (line 9 of @\color{codegreen}\cref{lst:clear_snapshot}@) */;
term vc2 = /* re-establish invariant
  (line 21 of @\color{codegreen}\cref{lst:clear_snapshot}@) */;
term vc3 = /* establish post-condition
  (line 23 of @\color{codegreen}\cref{lst:clear_snapshot}@) */;

thm proof1() {
    /* user-provided proof code for vc1 */
}
thm proof2() {
   /* user-provided proof code for vc2 */
}
thm proof3() {
   /* user-provided proof code for vc3 */
}

int main(void) {
    assert_prove(proof1(), vc1);
    assert_prove(proof2(), vc2);
    assert_prove(proof3(), vc3);
}
\end{lstlisting}
\caption{Residual Proof Program}\label{fig:example-residual-proof}
\end{subfigure}
\caption{Demonstration of \cstar{}'s workflow using the running example in \cref{sec:overview}.}\label{fig:example-workflow}
\end{figure}

\begin{remark}[Reversed role of verification-specific annotations]
Whereas in the deployment phase embedded proof-code blocks are ignored by the compiler, in the proof-checking phase, they play a central role.
% Assemble and slicing process
Here, the proof code assembled from embedded proof blocks becomes the main code for execution, while the annotated C code are sliced into segments and treated as serializable data, to be fed to the symbolic-execution engine interactively.
\end{remark}

\paragraph{Operational proof checking stage.}
In the second stage, the \cstar{} workflow executes the operational proof program obtained from the translation stage.
As discussed in \cref{sec:overview}, \cstar{} supports two styles of program verification, namely declarative and operational:
\begin{itemize}
    \item In the declarative style, the user asserts expected symbolic states at specific program points.
    The symbolic-execution engine uses the asserted symbolic state for further execution and produce \emph{verification conditions} as output.
    These verification conditions are gathered and will be proved \emph{in batch} later in the \emph{residual proof checking stage}.
    \item In the operational style, the user directly manipulates the symbolic state through proof-code blocks, which are executed interactively with the symbolic execution.
    In this way,  the operational proof checking phase is naturally \emph{real-time}: the proof-code blocks are executed \emph{interleaved} with symbolic execution, feeding the annotated program segments incrementally to the symbolic-execution engine.
\end{itemize}
More specifically, the \cstar{} workflow handles the execution of each proof block as follows.
First, the last annotated program segment---as some serializable data---is fed to the symbolic-execution engine.
Next, the proof-code block, which is normal C code, is executed in the proof-supporting runtime (see Remark \ref{rem:proof-supporting-runtime} below).
Finally, the current symbolic state is updated according to the execution results of the proof-code block: recall that a proof-code block should fetch the symbolic state by calling \lstinline|get_symbolic_state|, do transformations on it with proofs, and then call \lstinline|set_symbolic_state| at the end to update the symbolic state in the symbolic-execution engine.
The proof program shown in \cref{fig:example-operational-proof} implicitly calls these functions in the code of \lstinline|single_out_location|, i.e., \cref{lst:single_out_location}.

\begin{remark}[Proof-supporting runtime]
    \label{rem:proof-supporting-runtime}
    The proof programs are executed in \cstar{}'s proof-supporting runtime, which is the standard C runtime interfaced with the LCF-style proof kernel and the symbolic-execution engine.
    %
%    \begin{itemize}
        The LCF-style proof kernel implements the basic building blocks used in proof code.
        The symbolic-execution engine takes annotated C program segments as input
%        \footnote{It also requires an input file containing the signature of types and types of constants used in assertions.} 
        (via the built-in function \lstinline|feed_program_segment|), maintains the symbolic state of the current partial program internally (accessed and modified via the \lstinline|get_symbolic_state| and \lstinline|set_symbolic_state| built-in functions), and produces verification conditions as output when symbolic execution is completed.
        % Ideal primitives: get_symbolic_state, set_symbolic_state, feed_program_segment
%    \end{itemize}
\end{remark}

\paragraph{Residual proof checking stage.}
In the third stage, the \cstar{} compiler collects the output of the symbolic-execution engine and creates the \emph{residual proof program}, which is a \cstar{} program purely consisting of global proof-code blocks.
This program contains proof goals for every undischarged verification condition generated during symbolic execution, which are to be addressed in the \cstar{} proof environment, either by the programmers or with assistance from proof experts.
Those verification conditions arise from the declarative style of verification:
recall that at each assertion or invariant, it is obliged for \cstar{} users to prove the entailment from the maintained symbolic state (by the symbolic-execution engine) to the asserted state.
\cref{fig:example-residual-proof} shows the residual proof program for the running example in \cref{sec:overview}:
lines 1--6 are three verification conditions generated by the symbolic-execution engine,
lines 8--16 are user-provided proof code for the three verification conditions, respectively,
and the main function executes the proof code the check if they indeed prove the verification conditions.

\begin{remark}[\cstar{} developer's view]
    From the perspective of a developer, \cstar{} can be seen as an LCF-style higher-order-logic theorem prover embedded within C, tailored for C program verification.
    It relies on a trusted symbolic-execution engine that serves as an oracle, which automatically derives strongest post-conditions and generates verification conditions from C code segments annotated with separation-logic assertions and specifications.
    Successfully executing the two proof programs, i.e., operaational and residual, is thereby equivalent to verifying the Hoare triple of the entire program, which provides the correctness guarantee for the conformance with the program's specification.    
\end{remark}

\subsection{Interface for \texorpdfstring{\cstar{}}{C*} Users}
\label{sec:design:interface}

\begin{figure}[t]
\centering
\includegraphics[width=0.98\textwidth]{cstar-cheat-sheet.pdf}
\caption{A summary of the verification-specific interface that \cstar{} provides to users.}
\label{fig:cstar-cheat-sheet}
\end{figure}

As overviewed in \cref{sec:overview}, \cstar{} extends the C programming language with two categories of verification-specific syntactic constructs: (i) specifications and intermediate assertions, and (ii) proof-code blocks.
\cref{fig:cstar-cheat-sheet} summarizes these constructs, providing an accessible interface for \cstar{} users.
In this section, we explain our design of this interface and at the end exemplify \cstar{}'s extensibility in proof support using the implementation of \lstinline|local_apply| from \cstar{}'s standard proof-support library.

\paragraph{Verification-specific attributes}
\cstar{} introduces attributes \lstinline|[[require]]|, \lstinline|[[ensure]]|, \lstinline|[[parameter]]|, and \lstinline|[[argument]]|
concerning function specifications, \lstinline|[[assert]]| and \lstinline|[[invariant]]| for intermediate assertions within implementation code,
%(performing declarative verification, or serve as documentation),
as well as \lstinline|[[proof]]| for embedding proof-code blocks.
%(performing operational verification, or define common proof functions and theorems).
%

The attributes \lstinline|[[require]]| and \lstinline|[[ensure]]| specify a function's \emph{pre-condition} and \emph{post-condition}, respectively, both containing C expressions that evaluate to a \lstinline|term| value of object-logic type \lstinline|hprop|, i.e., a separation-logic predicate.
Both the pre- and post-condition can reference function parameters, and the post-condition can additionally reference the function's returned value using the preserved symbol \lstinline|__result|.
The attribute \lstinline|[[parameter(`var:type`)]]| introduces a \emph{ghost} parameter \lstinline|var| of object-logic type \lstinline|type| that denotes a universally quantified logical variable for the function specification.
Correspondingly, the \lstinline|[[argument(`var=value`)]]|---used before a function call---instantiates the ghost variable with an object-logic term.
\cref{fig:cstar-cheat-sheet} illustrates the usage of parameter and argument attributes using a C function \lstinline|reverse| that reverses a linked list in-place:
the parameter \lstinline|l| is a logic-level integer list that encodes the content of the linked list pointed to by \lstinline|p|,
where the (user-defined) separation-logic predicate \lstinline|ll_repr(p, l)| expresses such encoding.

Similar to \lstinline|[[require]]| and \lstinline|[[ensure]]|, the \lstinline|[[assert]]| and \lstinline|[[invariant]]| attributes take a C expression as input, which evaluates to a separation-logic predicate.
The \lstinline|[[assert]]| attribute inserts a static assertion about the symbolic state at a  program point, supporting a \emph{declarative} verification style: if non-trivial reasoning is required to prove that the maintained symbolic state entails the asserted state, a verification condition is generated by the symbolic-execution engine.
We will explain this workflow in detail in \cref{sec:workflow}.
After processing an assertion, the symbolic-execution engine will update the symbolic state accordingly.
The \lstinline|[[invariant]]| attribute also inserts an assertion but it asserts a symbolic state expected at the start of each loop iteration, hence its name \emph{invariant}.
Currently, only \lstinline|while| loops are supported, and using \lstinline|break| or \lstinline|continue| will lead the symbolic-execution engine to generate additional verification conditions for the additional control-flow paths.

The \lstinline|[[proof]]| attribute wraps a proof-code block.
Proof-code blocks may contain arbitrary C code, having access to the LCF-style proof kernel and the symbolic-execution engine.
There are two kinds of proof-code blocks: (i) \emph{local proof blocks}, used within implementation code (e.g., inside a function body), primarily for \emph{operational} verification and symbolic-state transformation, and (ii) \emph{global proof blocks}, used outside any implementation code, typically for defining common proof functions or theorems.
Typically, all code in a proof-support library (e.g., \lstinline|cstarlib.h|) is within global proof blocks.
Within a local proof block, the user can reference bindings declared in prior proof blocks in the same function body, as well as bindings declared in global proof blocks.

\paragraph{Specifications and intermediate assertions}
Inside the verification-specific attributes, \cstar{} provides a \emph{quotation} syntax (delimited by \lstinline|`...`|).
It allows the user to construct object-logic terms using concrete syntax representations and avoid the verbosity of calling term constructors explicitly.
Inside quotations, the user can use the \emph{anti-quotation} mechanism (escaped using \lstinline|${var:type}|) for splicing in computed sub-terms stored in program variables.
Together, these two syntax extensions offer a simple yet expressive way to build object-logic terms.
\cref{fig:cstar-cheat-sheet} summarizes the concrete syntax for separation-logic predicates and other frequently used object-logic terms and types.
There are a few unsual notational conventions, which arise from the LCF-style proof kernel employed by \cstar{}.
In the object logic, integer literals (i.e., terms of object-logic type \lstinline|integer|) take the form \lstinline|&n|, where \lstinline|n| is a natural number.
The logic-level representation of an address is an integer value, e.g., \lstinline|&"x"| denotes the address of the program variable named \lstinline|x|.
We reuse C's \lstinline|&&| and \lstinline!||! operators to encode logic-level conjunction an disjunction, respectively,
and use \lstinline|==>| for standard logical implication.
Separation-logic entailments are treated as propositions (i.e., terms of object-logic type \lstinline|bool|):
the binary operator \lstinline!|--! takes two separation-logic predicates \lstinline|hp1| and \lstinline|hp2| and produces a proposition \lstinline!hp1 |-- hp2!, whose meaning is that if a heap satisfies \lstinline|hp1|, then it also satisfies \lstinline|hp2|.

In \cstar{}, separation-logic predicates in specifications and intermediate assertions must adhere to a specific form to enable automated symbolic execution.
This specific form is known as the \emph{symbolic heaps}~\cite{DBLP:conf/aplas/BerdineCO05}, and has the following structure:
\begin{equation}\label{eq:symbolic-heap}
    \exists x_1, \ldots, x_k.\ (P_1 \land \cdots \land P_m) \land (Q_1 * \cdots * Q_n) \ , \tag{\textsc{SymHeap}}
\end{equation}
where $x_i$'s are existentially quantified logical variables, $\land$ denotes non-separating conjunction, i.e., standard logical conjunction, and $*$ represents separating conjunction.
The $P_i$'s are \emph{pure facts}---expressions in the form of \lstinline|pure(p)| that state properties about the global heap. 
The $Q_j$'s, known as \emph{spatial facts}, consist of either primitive maps-to predicates (i.e., \lstinline|data_at| or \lstinline|undef_data_at|) which are visible to the symbolic-execution engine, or user-defined predicates (e.g., \lstinline|array_at|) whose internal structure can be arbitrary and are opaque to the symbolic-execution engine.
These spatial facts represent separately-owned local fragments of the heap.
We often use the derived form \lstinline|fact(p) = pure(p) && emp| to describe pure properties. The derived form satisfies that \lstinline|pure(p) && H = fact(p) * H|.
This formulation allows symbolic heaps to be uniformly represented as separating conjunctions of pure and spatial facts, avoiding the need for non-separating conjunction.
% Free variables in the formula are implicitly universally quantified at the outer scope when proving entailments between symbolic heaps.

Before symbolically executing any primary program statement, the symbolic-execution engine verifies that the current symbolic state includes the necessary primitive maps-to predicates for all accessed memory locations.
Once this requirement is satisfied, the engine updates the symbolic state as needed, preserving the symbolic form, and possibly generates side conditions to guarantee safe execution, i.e., no runtime error or undefined behavior.

In addition to the primitive predicates and predicates provided in the standard library, \cstar{} users can derive and use their customized predicates.
For instance, the \lstinline|hiter| function in the standard proof-support library, defined in object-logic as \lstinline|hiter hps = fold_right (**) hps emp| using the higher-order function \lstinline|fold_right|, takes a list of separation-logic predicates \lstinline|hps| and returns their \emph{iterated separating conjunction}.
We will explain how to implement derived predicates later in this section.

\paragraph{Proof-code interface with symbolic execution}
In \cstar{}, a local proof-code block for performing operational verification retrieves the current symbolic state from the symbolic-execution engine by calling a built-in function \lstinline|get_symbolic_state()|.
For example, the initial symbolic state can be obtained with \lstinline|term pre_state = get_symbolic_state()|.
At the end of the proof-code block, the symbolic state can be updated using a call to \lstinline|set_symbolic_state(th)|, where \lstinline|th| is a theorem proving the separation-logic entailment from the current symbolic state (\lstinline|pre_state|) to a new state (\lstinline|new_state|).
This updated state \lstinline|new_state| is then set as the current symbolic state.
Recall \cref{lst:single_out_location} in \cref{sec:overview} for an example of using the interface to do local transformations on the symbolic state.

\paragraph{Proof-code interface with LCF-style proof kernel}
In an LCF-style proof environment like that in \cstar{}, two fundamental types are provided for logical reasoning: \lstinline|term|, representing terms in the object logic, and \lstinline|thm|, denoting proven theorems.
These types are treated as abstract types in \cstar{}, ensuring that users can only manipulate them through the library functions provided by the LCF proof kernel, forbidding direct access to internal data structures.
% Otherwise a theorem might be constructed without a formal proof.

As summarized in \cref{fig:cstar-cheat-sheet}, to work with \lstinline|term| values, the proof kernel offers a set of functions acting as constructors, destructors, discriminators, and equality checkers, among other utilities.
For \lstinline|thm| values, the kernel provides the primitive rules needed to prove theorems.
These rules encompass both separation-logic entailment rules and higher-order logic rules for general reasoning.
Additionally, functions for checking if a proof goal is achieved and for accessing the hypotheses and conclusion of a theorem are available.
%
%\diw{Such as ...}

The programmability of \cstar{}'s LCF-style proof kernel allows users to extend its functionality by defining customized derived rules or proof-search routines as C functions on top of the primitive proof rules.
Furthermore, besides the built-in types such as \lstinline|ctype| and \lstinline|hprop|, as well as standard functions like \lstinline|sizeof|, the kernel's \emph{definitional mechanism} enables users to define new (inductive) types, e.g., \lstinline|int_list| in \cref{fig:cstar-cheat-sheet}, as well as (recursive) functions, e.g., \lstinline|nth| in \cref{fig:cstar-cheat-sheet}.
Such definitional mechanism is also used to define new sepeartion-logic predicates, such as  \lstinline|hiter| mentioned earlier in this section.
% conservative extension by definition
This flexibility makes \cstar{} expressive for a wide range of verification needs.
%
%\cref{fig:cstar-cheat-sheet} exemplifies 

\paragraph{Extensible and programmable proof support}
With the programmability of LCF-style proof support, proof experts can develop custom proof libraries to simplify common proof patterns, offering high-level derived proof rules and collections of frequently used mathematical properties.

For example, a typical task in operational-style verification is justifying local transformations performed on the symbolic state.
By \emph{local transformations}, we mean picking out specific conjuncts from a symbolic heap, applying a proved-correct separation-logic entailment to these conjuncts, and leaving the rest of the symbolic heap unchanged.
Separation logic inherently supports such local transformations; however, using only primitive rules of separation logic---some of which are listed in \cref{fig:separation_logic_rules}--- requires manually lifting affected conjuncts through layers of separating conjunctions and specifying frames for each transformation.
This can be tedious and lead to proof code cluttered with structural manipulations, which detract from the intuitive reasoning process.

\begin{figure}[t]
\small
    \begin{mathpar}
    \inferrule[hsep-comm]{\ }{H_1*H_2 \reflectbox{$\vdash$}\mkern-5mu\vdash H_2*H_1}\and
    \inferrule[hsep-assoc]{\ }{(H_1*H_2)*H_3 \reflectbox{$\vdash$}\mkern-5mu\vdash H_1*(H_2*H_3)}\and
    \inferrule[hsep-cancel-right]{H_1 \vdash H_1'}
    {H_1 * H_2 \vdash H_1' * H_2}\and
    \inferrule[hexists-monotone]{\forall x.\ (H \vdash H')}{(\exists x.\ H) \vdash (\exists x.\ H')}
    \end{mathpar}
    \caption{Selected separation-logic rules for structural manipulations.}
    \label{fig:separation_logic_rules}
\end{figure}

To alleviate the need for manually performing such structural manipulations, we implemented a derived rule called \lstinline|local_apply| in our standard proof-support library \lstinline|cstarlib.h|.
Considers the simple case where only one conjunct is affected by the transformation, the automation process of \lstinline|local_apply| can be described in four steps as follows.
\begin{enumerate}[(i)]
  \item Repeatedly destruct existential binders in the symbolic heap of the form \eqref{eq:symbolic-heap}.
  \item Find the affected conjunct and lift it to the far-left side of the symbolic heap by using the \textsc{hsep-comm} and \textsc{hsep-assoc} rules repeatedly.
  \item Apply the \textsc{hsep-cancel-right} rule with other conjuncts to the right as the frame.
  \item Repeatedly add back existential binders using the \textsc{hexists-monotone} rule.
\end{enumerate}
As a concrete code example, we present the proof function \lstinline|sep_lift_one| for performing the second step in \cref{lst:sep_lift_one}.
It assumes the input symbolic heap \lstinline|septerm| is a right-associated separating conjunction.
It first calls the derived rule \lstinline|hsep_move|, getting a generalized equality theorem \lstinline|lift_to_left| (line 4) for moving the target conjunct out to the left for one layer when it is in the left position of the inner symbolic heap, in one step.
It then tries to find the target conjunct recursively:
\begin{enumerate}[(i)]
  \item If the target conjunct is never found, it returns \lstinline|NULL|.
  \item If the target conjunct is at the far-right position, it rewrites it using the \textsc{hsep-comm} rule.
  \item Otherwise, it uses the equality theorem \lstinline|lift_to_left| to move the target conjunct out to the left for one layer.
  The lifting steps work in bottom-up way during unwinding the recursive calls.
\end{enumerate}

\begin{figure}[t]
  \centering
  \begin{lstlisting}[language=cstar, caption={\cstar{} code of \lstinline|sep_lift_one|.}, label={lst:sep_lift_one}]
thm sep_lift_one(term target, term septerm)
    /* assume target is primitive and septerm is a right-associated symbolic heap */
{
    thm lift_to_left = hsep_move(target);
    /* @$\color{codegreen}\forall$@ (hp1:hprop) (hp2:hprop).
         hp1 ** ${target:hprop} ** hp2 -|- ${target:hprop} ** hp1 ** hp2 */
         
    if (is_sep(septerm)) {
        term l = left_of_sep(septerm), r = right_of_sep(septerm);
        if (equals_term(target, l)) { return rewrite(lift_to_left, septerm); } 
        else {
            if (is_sep(r)) { 
                thm step1 = rewrite(sep_lift_one(target, r), septerm); 
                thm step2 = rewrite(lift_to_left, consequent(conclusion(step1)));
                return trans(step1, step2);
            } else if (equals_term(target, r)) 
                return rewrite(symm(hsep_comm(target)), septerm);
        }
    } else if (equals_term(target, septerm)) { return refl(septerm); }
    return NULL; /* fail if target isn't found */
}
\end{lstlisting}
\end{figure}

\section{Implementation and Evaluation}
\label{sec:evaluation}

In this section, we describe our prototype implementation of \cstar{} and our evaluation of it.
In \cref{sec:implementation}, we discuss some aspects of our prototype \cstar{} implementation diverged from the core design in \cref{sec:design}.
In \cref{sec:benchmarks}, we present an empirical evaluation of our prototype implementation on a suite of C benchmark programs and report some interesting findings on using \cstar{} for program verification.

\subsection{Implementation Notes}\label{sec:implementation}
%In the previous section, we discuss the ideal workflow for \cstar{} in \cref{sec:workflow}, and illustrate it in \cref{fig:cstar-conceptual-workflow}.
%
Following the workflow described in \cref{sec:workflow}, our implementation consists of three main components: the \cstar{} compiler, the LCF-style proof kernel, and the symbolic-execution engine.

%along with some of the design decisions and workarounds we applied in the current implementation.
%
% For the three main components, we implement the \cstar{} compiler in OCaml, reuse the LCF-style proof kernel of HOL Light, and rely on an external symbolic execution engine.
%

\paragraph{Implementing the \cstar{} compiler}
The \cstar{} compiler, implemented in OCaml, processes C code with \cstar{}-specific attributes, managing syntax extensions (quotation and anti-quotation) and translating them to invocations of term-parsing functions and substitution primitives.
The compiler also assembles code in the proof blocks to form a proof program that executes in the proof-supporting runtime.
Note that the proof program is a C program.
Specifically, global proof blocks are moved to the beginning of the generated C program, and each function in the implementation code creates a proof function, with local proof blocks appended in the order of their appearance.
It also aligns the concrete annotation syntax (and separation-logic assertion syntax) used in \cstar{} with the external symbolic execution-engine.
%some of which are shown in \cref{tab:annotated-assertions}.

%\begin{table}[t]
%    \centering\small
%    \caption{Corresponding annotations for \cstar{} attributes and functions in the symbolic-execution engine.}
%    \begin{threeparttable}
%        \begin{tabular}{|l|l|l|}
%            \hline
%            \textbf{\cstar{} Assertions/Functions}        & \textbf{Supported Annotations}                        & \textbf{Meaning}              \\
%            \hline
%            \lstinline|[[parameter(`var:type`,...)]]| & \lstinline|/*@ With (var:type)... */| & ghost parameters \\
%            \lstinline|[[argument(`var=value`,...)]]| & \lstinline|/*@ Where (var=value)... */| & ghost arguments    \\
%            \lstinline|[[require(hp)]]|                & \lstinline|/*@ Require hp */|           & precondition                  \\
%            \lstinline|[[ensure(hp)]]|                 & \lstinline|/*@ Ensure hp */|            & postcondition                 \\
%            \lstinline|[[assert(hp)]]|                 & \lstinline|/*@ Assert hp */|            & assertion with VC generated \\
%            \lstinline|[[invariant(hp)]]|              & \lstinline|/*@ Assert Inv hp */|        & loop invariant                \\
%            \lstinline|get_symbolic_state()|           & \lstinline|/*@ Get id */|               & retrieve the symbolic state   \\
%            \lstinline|set_symbolic_state(th)|         & \lstinline|/*@ Assume new_st */|            & assertion w/o VC generated \\
%            \hline
%        \end{tabular}
%    \end{threeparttable}
%    \label{tab:annotated-assertions}
%\end{table}

\paragraph{Reusing HOL Light proof kernel}
In the implementation of \cstar{}, we reuse the LCF-style proof kernel of the HOL Light prover~\cite{DBLP:conf/tphol/Harrison09a}, a minimal implementation of higher-order logic in OCaml.
This avoids the need to build a new LCF-style kernel from scratch in C, while leveraging the extensive libraries available in HOL Light for mathematical reasoning.
%
% We wrap and expose a first-order interface to manipulate the object-logic terms and theorems in C (detailed in \cref{sec:design}).
%
%This setup uses a client-server architecture, where the HOL Light proof kernel runs in a separate process.
%%
%Proof programs interface with the kernel using RPC (Remote Procedure Call) mechanisms, which can potentially be a bottleneck for performance.
%
To support separation-logic entailment proofs needed in program verification, we axiomatize a separation logic theory in HOL Light with a concrete memory model in mind, interpreting the heap as a finite mapping from addresses to bytes and treating integers and pointers the same in higher-order logic.

\paragraph{Interfacing with the symbolic execution engine}

In the ideal workflow illustrated in \cref{fig:cstar-conceptual-workflow}, proof programs communicate with the symbolic-execution engine via functions like \lstinline|get_symbolic_state()|, \lstinline|set_symbolic_state(th)|, and \lstinline|feed_program_segment(prog)|.
However, we currently lack access to the internal states of the external symbolic-execution engine, making it challenging to implement this interactive workflow directly.
Consequently, we currently rely on the annotations that the symbolic-execution engine supports for communication.
%
%shown in \cref{tab:annotated-assertions}: \lstinline|/*@ Get id */| for getting the symbolic state at a specific program point and bind it to the \lstinline|id| item in the output, and \lstinline|/*@ Assume new_st */| for setting the symbolic state to a specific assertion without generating a verification condition.\footnote{Assuming the current symbolic state is \lstinline|pre_st|, the theorem \lstinline|th| should prove \lstinline$pre_st |-- new_st$.}
%
To simulate the interleaving execution pattern in the ideal workflow, we need to run the symbolic-execution engine twice for each proof block: once for getting the symbolic state and once for setting it after running the proof code.
This is done manually for now.

\subsection{Empirical Evaluation}
\label{sec:benchmarks}

To evaluate the effectiveness of our prototype implementation of \cstar{},
we selected a suite of small C programs as benchmarks and verified their functional correctness using \cstar{}.
Most examples are adapted from the VeriFast repository~\cite{DBLP:conf/nfm/JacobsSPVPP11}, while the buddy allocator example is drawn from CN~\cite{DBLP:journals/pacmpl/PulteMSMSK23}.
Some additional examples were crafted manually to test \cstar{}'s handling of complex control-flow structures.
This benchmark allows us to (i) test the functionality of the \cstar{} toolchain, including its frontend parser, proof-supporting runtime, and the translation phase, (ii) assess the expressiveness of \cstar{}'s reasoning capabilities, and (iii) evaluate the usability of \cstar{}'s hybrid operational and declarative proof approach.

\begin{table}[t]
    \centering\small
    \caption{
        Evaluation of \cstar{}. ``\textbf{Impl}'' is short for ``Implementation Code.''
        ``\textbf{PB}'' is short for ``Proof Block.''
        ``\textbf{VC}'' is short for ``Verification Condition.''
        ``\textbf{Spec}'' is short for ``Specficiation.''
        ``\textbf{Assert}'' is short for ``Assertion.''
    }
    \label{fig:evaluation}
    \begin{tabular}{clccccc}
        \toprule
        \textbf{Class}                & \textbf{Name}                   & \textbf{\#Line of Impl} & \textbf{\#PB} & \textbf{\#VC} & \textbf{\#Line of Proof}  & \textbf{\#Line of Spec/Assert} \\ \midrule\addlinespace[0.5pt]\midrule
        \multirow{3}{*}{\#1} & address\_of\_local     & 32   &     4          &   0   &       14        &             10          \\ \cline{2-7}
                             & globals                & 18   &     3          &   0   &       20        &              4          \\ \cline{2-7}
                             & swap                   & 15   &     0          &   0   &        0        &              7          \\ \midrule
        \multirow{3}{*}{\#2} & multi\_branch          & 24   &    10          &   0   &      102        &             25          \\ \cline{2-7}
                             & mutually\_recursive    & 17   &     2          &   6   &      69        &             13          \\ \cline{2-7}
                             & no\_return               & 15   &     1          &   0   &        4        &              9          \\ \midrule
        \#3                  & malloc\_free           & 9    &     1          &   0   &        9        &              8          \\ \midrule
        \multirow{3}{*}{\#4} & clear                  &  9   &     7          &   0   &       120       &            11           \\ \cline{2-7}
                             & forall                 & 10   &     7          &   0   &      153        &             13          \\ \cline{2-7}
                             & reverse                & 18   &     7          &   1   &      375        &             58          \\ \midrule
        \#5                  & attach\_page & 57   &     6          &   3   &     1616        &            451          \\ \bottomrule
    \end{tabular}
\end{table}

A complete list of the benchmark is shown in \cref{fig:evaluation}
and the source code of all benchmark programs is included in the Supplementary Material.
%
%\cyy{TODO: explain the information in the table.}
% 
We chose these programs to encompass a broad spectrum of reasoning patterns, including shared memory access (\#1), control-flow constructs (\#2), dynamic memory management and interaction with external functions (\#3), complex model-level reasoning (\#4), and a real-world case study (\#5).
\cref{fig:evaluation} presents statistics regarding the code size of each benchmark program.
The column ``\textbf{\#Line of Impl}'' lists the number of lines of implementation code.
The total number of lines in each benchmark program is significantly larger due to the inclusion of proof code, whose statistics is given in the column ``\textbf{\#Line of Proof},'' as well as specifications and assertions, whose statistics is given in the column ``\textbf{\#Line of Spec/Assertion}.''
We also include (i) the number of proof blocks for the operational style of verification and (ii) the number of verification conditions for the declarative style of verification.
%  demonstrating our support for both methodologies.
%
%We present evaluation highlights below.

%\cyy{TODO: in the uploaded cstar files, most [[cstar::assert]] are for documentation, not for declarative verification.}

\paragraph{Coverage of C language features}
The evaluation demonstrates \cstar{}'s support for core C features, especially those that create flexible aliasing patterns and complex control flow structures:
\begin{itemize}
    \item Control-flow constructs, including multiple branching (\lstinline|if...else if...|), (mutually) recursive functions, \lstinline|break| and \lstinline|continue|, and (early) \lstinline|return|.
    Benchmark programs \lstinline|address_of_local|, \lstinline|multi_branch|, \lstinline|mutually_recursive|, and \lstinline|no_return| use some of these constructs.
    \item Shared memory access, covering global variables, arrays, addressable local variables, and (multi-level) pointer indirections. Benchmark programs \lstinline|address_of_local|, \lstinline|globals|, and \lstinline|swap| make use of shared memory access.
    \item Dynamic memory management and interaction with (formally specified) external functions, tested via \lstinline|malloc| and \lstinline|free|. The benchmark program \lstinline|malloc_free| demonstrate those features. 
\end{itemize}
Currently, our prototype implementation does not support \lstinline|switch| statement, \lstinline|goto|, or other looping constructs (i.e., \lstinline|for| and \lstinline|do while|).
We leave supporting those features for future work.

%\paragraph{Common systems programming idioms}
%%
%Some examples demonstrate \cstar{}'s handling of low-level systems programming idioms, such as pointer arithmetic and type casting.
%\cyy{filename and description here}

\paragraph{Complex logical reasoning}
The benchmarks also illustrate \cstar{}'s capability for performing complex logic-level reasoning.
The expressiveness of higher-order logic used in \cstar{} enables users to define functional models (as inductive data types, e.g., lists or trees) and also (well-founded) recursive functions that operate on these models (e.g., reversing a list), using high-level definitional mechanisms.
Users can also define recursive representation predicates to link the entry points of concrete memory structures to their functional models, a technique typical of separation-logic-based program reasoning~\cite{DBLP:conf/cpp/Chargueraud16}.
%
%\cyy{TODO: add more details in examples here}
%
In several instances within our benchmark, we leveraged the pre-existing proof libraries of HOL Light, thereby reducing the effort required for model-level reasoning.
Nonetheless, for the benchmark program \lstinline|reverse|, we proved four logic-level reasoning lemmas and two ownership-related reasoning lemmas, which are reusable for reasoning about linked lists.

The proof code in those benchmark programs extensively uses \cstar{}'s standard proof-support library.
In addition to \lstinline|local_apply| described in \cref{sec:design:interface}, our proof-support library includes other reusable derived rules.
For example,
\lstinline|sep_normalize(t)| transforms a heap proposition into a canonical form, 
\lstinline|sep_lift(l,t)| lifts a sub-part of the heap proposition \lstinline|t| to the far-left side, generalizing the \lstinline|sep_lift_one| function in \cref{lst:sep_lift_one},
and \lstinline|sep_reorder(t1,t2)| verifies if two heap propositions are reorderings of each other (modulo $\alpha$-renaming of bound variables).

\paragraph{A real-world case study: buddy allocator}
Inspired by CN~\cite{DBLP:journals/pacmpl/PulteMSMSK23},
we applied \cstar{} to a more challenging real-world case study: the attach function of the buddy allocator used in pKVM \cite{misc:pKVM}.
A buddy allocator manages memory in blocks of size $2^o \times 4$ KB, where $o \in {0, 1, \ldots, \mathit{max\_order}-1}$ denotes the order of the block.
Each block is aligned according to its size, maintaining an invariant about the alignment for all blocks.

Two blocks are called \emph{buddies} if they (i) are adjacent, (ii) have the same order, and (iii) can be merged into a larger block of the next order while preserving alignment.
Allocatable memory is divided into pools, each representing a contiguous range of pages.
Every pool maintains a doubly-linked list of free blocks for each order, and the allocator searches these lists for a free block of the required size during memory allocation.
Readers may refer to~\cite{DBLP:journals/pacmpl/PulteMSMSK23} for further details on the data structures and helper functions used in this case study.

We verified the \lstinline|attach_page| function, shown in \cref{lst:attach_function_body}, from the implementation of the buddy allocator.
This function operates by receiving a released block, identifying any adjacent free buddy block in the pool, and merging them to form a larger free block.
This merging process continues iteratively until no more free buddies are found or the maximum order is reached.
The resulting block is then added back to the pool.
The loop invariant of the while loop in the implementation code is shown in \cref{lst:attach_invariant}.
An interesting finding during verification was that the specifications in CN~\cite{DBLP:journals/pacmpl/PulteMSMSK23} were not sufficient to guarantee that all free blocks are present in a doubly-linked list. Despite this, we adhered to these weaker specifications for simplicity in our verification efforts.

\begin{figure}[t]
\centering
\begin{subfigure}[b]{0.47\textwidth}
\begin{lstlisting}[
%    caption={The \lstinline|attach_page| function.},
%    label={lst:attach_function_body},
    xleftmargin=0pt,
    numbers=none,
]
struct hyp_page *__hyp_vmemmap;
static void attach_page(
  struct hyp_pool *pool, struct hyp_page *pg
) {
  struct hyp_page *buddy = NULL;
  u8 order = pg->order;
  pg->order = (u8)HYP_NO_ORDER;
  u8 max_order_ = pool->max_order;
  memset_page_zero(pg,order);
  buddy = __find_buddy_avail(pool,pg,order);
  while ((order + 1) < max_order_ && 
         buddy != NULL) {
    page_remove_from_list_pool(pool,buddy);
    buddy->order = (u8)HYP_NO_ORDER;
    pg = min(pg, buddy);
    order = order + 1;
    buddy = __find_buddy_avail(pool,pg,order);
  }
  pg->order = order;
  page_add_to_list_pool(pool,pg,order);
}
\end{lstlisting}
\caption{The implementation code.}\label{lst:attach_function_body}
\end{subfigure}
\begin{subfigure}[b]{0.52\textwidth}
\begin{lstlisting}[
%    caption={The loop invariant of \lstinline|attach_page| function.},
%    label={lst:attach_invariant},
    xleftmargin=0pt,
    numbers=none,
]
[[invariant(`
exists buddy_v bi inv_l inv_dl inv_hl i order_v pg_v.
  data_at(&"max_order", Tuchar, &max_order) **
  data_at(&"order", Tuchar, &order_v) **
  data_at(&"pg", Tptr, pg_v) **
  data_at(&"buddy", Tptr, buddy_v) **
  data_at(&"pool", Tptr, pool_pre) **
  data_at(&"__hyp_vmemmap", Tptr, vmemmap) **
  (dlist_head_repr pool_pre 0 max_order inv_hl) **
  (free_area_repr
    (is_free_1st inv_l) start end inv_l) **
  (free_area_head_repr
    (is_free_1st inv_l) start end inv_dl) **
  (store_pageinfo_array vmemmap start end inv_l) **
  (store_zero_array
    (i2vaddr i) 0 (PAGE_SIZE * (2 EXP order_v)) 
    (PAGE_SIZE * (2 EXP order_v))) **
  ${other_facts_and_representation_predicates:hprop}
`)]]
\end{lstlisting}
\caption{The invariant of the loop.}\label{lst:attach_invariant}
\end{subfigure}
\caption{The \lstinline|attach_page| function.}
\end{figure}

\paragraph{Experience report}

The experience of two undergraduate students in using \cstar{} for benchmark evaluation reveals several usability issues of the current prototype implementation:
\begin{itemize}
    \item \emph{IDE support.} The lack of an IDE that shows symbolic states alongside code was a major pain point.
    Currently, users must run the symbolic execution engine manually and inspect symbolic states from its lengthy output, which interrupts the workflow.
    Developing an IDE for \cstar{}, e.g., as an editor plugin, is left for future work.
%    We are actively developing an IDE plugin that integrates a symbolic-state side-by-side viewer to address this issue.
    %
    \item \emph{Proof automation.} \cstar{} lacks automation for discharging trivial facts, making simple proofs time-consuming.
    This is partly due to the absence of solver-aided proof automation, heavily relied on by tools like CN~\cite{DBLP:journals/pacmpl/PulteMSMSK23} and VeriFast~\cite{DBLP:conf/nfm/JacobsSPVPP11}.
    Looking forward, we plan to provide \cstar{} an interface to encode and delegate proof obligations to external automated theorem provers or frameworks such as Z3~\cite{DBLP:conf/tacas/MouraB08} and Why3~\cite{DBLP:conf/esop/FilliatreP13}.
%    when trusted code base is not of primary concern.
    %
    \item \emph{Proof-support library.} Writing separation-logic entailment proofs in \cstar{} currently requires considerable boilerplate code, leading to long and repetitive proof code.
    This issue arises because \cstar{} lacks a rich set of derived rules for handling separation-logic reasoning, unlike mature frameworks such as Iris~\cite{DBLP:conf/popl/KrebbersTB17}, VST~\cite{DBLP:journals/jar/CaoBGDA18}, or CFML~\cite{DBLP:journals/pacmpl/Chargueraud20}.
    In our future work, we expect that expanding \cstar{}'s proof support libraries with more derived rules could improve the conciseness of proof code and developer productivity.
    %
%    The development of such derived rules are discussed in \cref{sec:design:interface}.
\end{itemize}

\section{Related Work}
\label{sec:relatedwork}

\paragraph{\textbf{Live Verification framework}}
The Live Verification framework~\cite{DBLP:journals/pacmpl/GruetterFC24} is a recently proposed framework with a similar goal of enabling its users to verify their low-level code as they write it.
The framework is embedded in the Coq proof assistant and provides real-time display of the symbolic state at the cursor position in the goal panel.
After a function has been given a prototype with formal specifications, users develop the function body incrementally by either writing the next line of implementation code,
%\footnote{Internally, these are Coq \emph{notations} for weakest-precondition rule applications with \emph{existential meta-variables}}
or writing Ltac proof scripts to shift the view on the symbolic state or discharge generated side conditions.
When this derivation process is finished, a correctness proof is produced alongside the assembled implementation code.
With some clever tricks, these Ltac source files can also be viewed as ordinary C code (with Ltac proof scripts in comments) and compiled directly with C compilers.

A key difference between \cstar{} and the Live Verification framework is \cstar{}'s focus on accessibility for conventional programmers.
In the Live Verification framework, proof development and customization of proof automation require proficiency in Coq's Ltac tactic language, which diverges from the imperative programming experience familiar to programmers.
In contrast, \cstar{} allows proof code to be written directly in the same language as the implementation code, making it more approachable for conventional programmers.
%
%The Live Verification framework also supports specifying loop invariants as incremental ``diffs'' of the symbolic state before the loop's start, potentially improving user experience and proof maintainability.
%%
%In \cstar{}, users can achieve similar functionality by writing proof blocks to perform fine-grained, transformation-based reasoning on the symbolic state.
%
% CStar also supports first-class assertions that can be another method to alleviate the burden of writing down the full assertion by hand.

\paragraph{\textbf{VeriFast}}
VeriFast~\cite{DBLP:conf/nfm/JacobsSPVPP11} is a state-of-the-art symbolic execution and separation logic-based automated verification tool for C and Java.
It has a custom specification language that allows users to define inductive data types, structurally recursive functions, and recursive representation predicates.
VeriFast emphasizes predictable automation: during symbolic execution, users manually unfold and fold predicates using the proof commands \lstinline[language=C]|open| and \lstinline[language=C]|close|.
A restricted form of existential quantification is supported in the form of pattern matching, and reasoning on first-order values are delegated to the SMT solver.
When inductive reasoning is required, it supports user-written ghost lemma functions, which are verified like the implementation code but require proof of termination and must be observationally pure.
VeriFast can handle a substantial subset of C features.

The primary distinction between \cstar{} and VeriFast lies in the extensibility of their proof support.
In VeriFast, proof support is limited to a fixed set of built-in ghost statements and basic induction capabilities using lemma functions.
On the other hand, \cstar{} enables users to develop custom proof rules and automation functions, offering greater flexibility and expressiveness for complex verification tasks.
Also, this extensibility allows experts to create high-level reasoning abstractions that are accessible to  programmers.

%\paragraph{\textbf{RefinedC}}
%%
%RefinedC~\cite{DBLP:conf/pldi/SammlerLKMD021} is a customizable ownership and refinement type system for C, offering syntax-directed proof automation with a foundational soundness guarantee.
%%
%Built on the semantic typing approach, RefinedC interprets types as separation logic predicates within the Iris separation logic framework, treating typing rules as lemmas that proof experts can add and prove within Coq.
%%
%These rules are collected as a logic program in RefinedC's separation logic programming language, Lithium, enabling non-backtracking syntax-directed proof automation during type checking.
%%
%The system is foundationally sound with respect to its custom C semantics, Caesium.
%%
%When the specially-designed typing rules fail to capture the ownership pattern applied in the code, or the automation tactics fall short to solve the pure side conditions generated during type checking, proof experts need to step in to fill the gap by embedding proof scripts in Ltac or modify the typing rules.

\paragraph{CN}
CN~\cite{DBLP:journals/pacmpl/PulteMSMSK23} is an ownership and refinement type system for C, targeting the verification of real-world systems software.
CN aims for predictable proof automation, employing the Liquid types~\cite{DBLP:conf/pldi/RondonKJ08} approach for decidable automation using an SMT backend, with heuristics for instantiating quantifiers.
It supports sound ownership reasoning at the type level using idea similar to capabilities~\cite{DBLP:journals/fuin/AhmedFM07}, split the type of a heap fragment into a linear capability type and an unrestricted pointer type for flexible aliasing commonly found in real-world  code.
Additionally, CN is grounded on a realistic semantics, Cerberus~\cite{DBLP:conf/pldi/MemarianMLNCWS16}, which accurately models a large fragment of ISO C.

%\paragraph{\textbf{Low*}}
%%
%Low*~\cite{DBLP:journals/pacmpl/ProtzenkoZRRWBD17} is an embeded domain-specific language for low-level C code verification in the F* language~\cite{DBLP:conf/popl/SwamyHKRDFBFSKZ16}.
%%
%As a first-order subset of F*, Low* uses a C-like memory model and can be transpiled to C code through the KaRaMeL compiler.
%%
%Because Low* is embedded in F*, it inherits F*'s support for automated dependent-refinement types, which is highly expressive.
%%
%It has been successfully applied by experts in many projects~\cite{DBLP:conf/sp/ProtzenkoPFHPBB20, DBLP:conf/pldi/SwamyRRSNMVTCG22}.
%
%\paragraph{\textbf{VST-A}}
%%
%VST-A~\cite{DBLP:journals/pacmpl/ZhouQWAC24} is a foundational assertion-based verifier for C, built on top of the VST and CompCert toolchains.
%%
%By proving a soundness theorem for splitting the verification of an annotated program into the verification of straightline Hoare triples, VST-A enables the assertion-based verification workflow instead of the interactive experience of VST.
\section{Conclusion}
\label{sec:conclusion}

In this paper, we presented \cstar{}, a new system and language design for verified programming in C.
%
%\cstar{} is designed to be accessible for conventional C programmers with minimal learning overhead while enabling proof experts to expand its proof capabilities through reusable proof libraries.
%
\cstar{} provides three key features:
(i) language-level integration supporting both declarative and operational styles of verification,
(ii) comprehensive reasoning capabilities within an expressive logic using C's programming paradigm, and
(iii) support for real-time verification.
It builds upon the established techniques of separation logic-based symbolic execution for modular program reasoning, as well as the LCF-style approach to programming proofs.
We implemented a prototype of \cstar{} and evaluated its effectiveness by developing verified C programs using a suite of benchmark programs.
In the future, we plan to develop an IDE for \cstar{} to enable interactive program verification,
interface \cstar{} with solved-aided proof automation to reduce proof efforts,
and develop more proof-support libraries for \cstar{}.

\bibliographystyle{ACM-Reference-Format}
\bibliography{verif,db}

% \newpage
% \appendix

%\input{proof}

\end{document}